\documentclass{ws-procs9x6}

\def\er#1#2{\relax\ifmmode{}^{+#1}_{-#2}\else$^{+#1}_{-#2}$\fi}
\def\slashchar#1{\setbox0=\hbox{$#1$}
   \dimen0=\wd0 \setbox1=\hbox{/} \dimen1=\wd1
   \ifdim\dimen0>\dimen1 \rlap{\hbox to \dimen0{\hfil/\hfil}} #1
   \else  \rlap{\hbox to \dimen1{\hfil$#1$\hfil}} / \fi}

\begin{document}

\title{Chiral Symmetry and $s-$wave Low-Lying Meson-Baryon
Resonances\footnote{\uppercase{T}his work is supported by
\uppercase{DGES} under contract \uppercase{BFM}2000-1326, by the
\uppercase{J}unta de \uppercase{A}ndaluc\'\i a, and \uppercase{DGI} and
\uppercase{FEDER} funds, under contract \uppercase{BFM}2002-03218.}}

\author{J. Nieves, C. Garc\'{\i}a--Recio, E. Ruiz Arriola} 

\address{Departamento de Fisica Moderna, Universidad de Granada,
E-18071 Granada, Spain }

\author{M. J. Vicente Vacas}

\address{Departamento de Fisica Te\'orica and IFIC, 
Centro mixto Universidad de  Valencia-CSIC, 
Aptd. 22085, E-46071 Valencia, Spain}  

\maketitle

\abstracts{
  The $s-$wave meson-baryon scattering is analyzed for the
  isospin-strangeness $I=1/2, S=0$ and $I=0,S=-1$ sectors, in a
  Bethe-Salpeter coupled channel formalism incorporating Chiral
  Symmetry. For both sectors, four channels have been considered: $\pi
  N$, $\eta N$, $K \Lambda$, $K \Sigma$ and $\pi \Sigma$, $\bar K N$,
  $\eta \Lambda$, $K \Xi$, respectively. The needed two particle
  irreducible matrix amplitudes are taken from lowest order Chiral
  Perturbation Theory in a relativistic formalism. There appear
  undetermined low energy constants, as a consequence of the
  renormalization of the amplitudes, which are obtained from fits to
  the available data: elastic $\pi N $ phase-shifts, $\pi^- p \to \eta
  n$ and $\pi^- p \to K^0 \Lambda$ cross sections and to
  $\pi\Sigma\to\pi\Sigma$ mass-spectrum, the elastic $\bar K N \to
  \bar K N$ and $ \bar K N\to \pi \Sigma$ $t$--matrices and to the $
  K^- p \to \eta \Lambda$ cross section data. The position and
  residues of the complex poles in the second Riemann sheet of the
  scattering amplitude determine masses, widths and branching ratios
  of the $S_{11}-$ $N$(1535) and $-N$(1650) and $S_{01}-$
  $\Lambda$(1405) and $-\Lambda$(1670) resonances, in reasonable
  agreement with experiment. A good overall description of data, from
   threshold up to around 2 GeV is achieved despite the fact
  that three-body channels have not been explicitly included.}


\section{Introduction}

Baryon resonances are outstanding features in elastic and inelastic
meson-baryon scattering and signal the onset of non-perturbative
physics.  At low and intermediate energies, renouncing to find out a
picture of the hadron as a valence quark bound state, it seems
appropriate to start considering the hadrons as the relevant degrees
of freedom where chiral symmetry not only proves helpful to restrict
the type of interactions between mesons and baryons, but also provides
an indirect link to the underlying QCD\cite{Pich95}.  Then, resonances
manifest themselves as poles of the scattering amplitude in a certain
Riemann sheet in the complex energy plane. Chiral Perturbation Theory
(ChPT) leads to amplitudes that, though exactly satisfy crossing
symmetry, are based on a perturbative expansion of a finite amount of
Feynman diagrams. This complies with unitarity order by order in the
expansion, but fails to fulfill exact unitarity of the scattering
amplitude. Thus, some non-perturbative resummation should be
supplemented to restore exact unitarity and, hopefully, to accommodate
resonances. Several unitarization methods have been suggested in the
literature to describe the meson-baryon dynamics: Inverse Amplitude
Method\cite{pn00,ej00c}, dispersion
relations\cite{JE01,Meissner:1999vr}, Lippmann-Schwinger equation 
 and Bethe-Salpeter Equation (BSE)\cite{KSW95}$^-$\cite{LK02}.

In this talk, we will study the $s-$wave meson-baryon scattering for
the strangeness $S=0$ and $-1$ and isospin $I=1/2$ and 0 respectively,
sectors. These reactions provide good examples of the need of
unitarization methods, e.g. in the $S=-1$ channel Heavy Baryon ChPT to
one loop fails completely already at threshold\cite{Ka01}, due to the
the nearby subthreshold $\Lambda (1405) $-resonance.  It is important
to realize from the very beginning that the dynamics of these
reactions requires the use of a coupled channel formalism, thus we
solve the BSE, in coupled channels, with a kernel determined by the
lowest order ChPT. After renormalization of the amplitude, a finite
number of phenomenological constants, which encode the detailed
underlying short-distance dynamics are required. Since, there is no
possibility to compare with ChPT beyond leading order, we have been
forced to fit the unknown Low Energy Constants (LEC's) to data. Thus,
at the end of the day, we come up with amplitudes which satisfy
exactly coupled channel unitarity, but violate crossing symmetry at
subleading order. Restoring crossing symmetry at some order, would
reduce the number of undetermined LEC's\cite{EJ99}, but, as already
noted, since the coupled channel ChPT one-loop amplitude for
meson-baryon scattering is not known, such program becomes quite
difficult to implement. Thus, we will pay special attention to
statistical correlations between the fitted LEC's to shed light on
this issue. This talk is based on the results of Refs.~\refcite{JE01b}
($S=0$) and~\refcite{prd} ($S=-1$), where the reader can find further
details.
 
\section{Theoretical framework and Results} 
\begin{figure}[ht]
   \epsfxsize = 9cm
   \centerline{\epsfbox{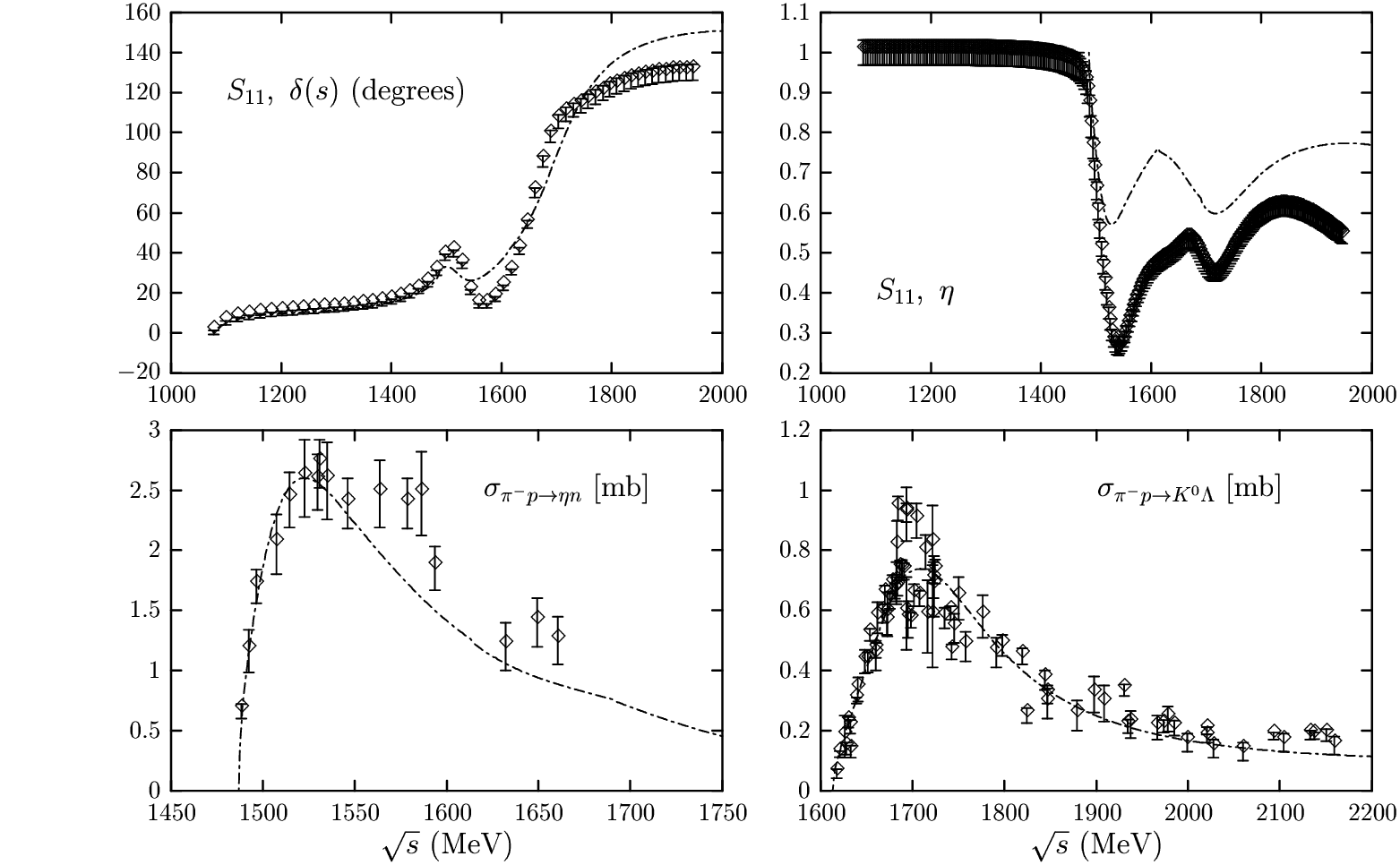}}
\vspace{-6cm}
\resizebox{0.9\textwidth}{!}{%
\includegraphics{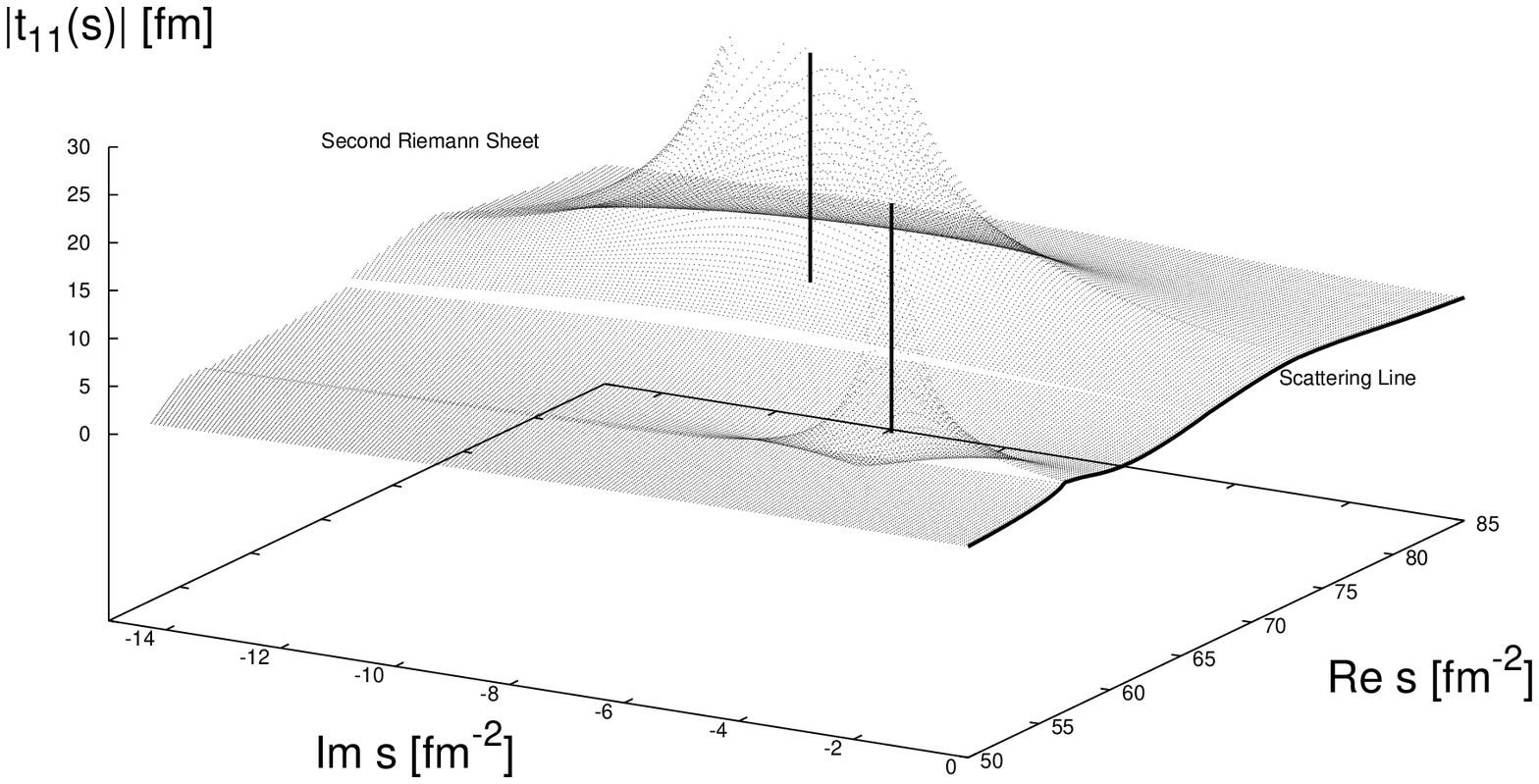}} 
\caption{ Best fit results for the Bethe-Salpeter equation in the
$S=0$, $I=1/2$ meson-baryon sector.  Upper panels: $\pi N $ $S_{11}$
phase shifts (left) and inelasticities (right) as a function of
$\sqrt{s}$. Note that the discrepancies
between theory and data are mainly due to the effect of three body
channel not included in the model. Middle panels: $\pi N \to \eta N $
(left) and $\pi N \to K \Lambda $ (right) cross sections. Lower panel:
Modulus of the $\pi N \to \pi N$ element of the scattering amplitude
in the $s-$complex plane (fourth quadrant of the second Riemann
sheet). Vertical lines indicate the position of the poles.  The two
observed poles are identified to be the $S_{11}-$ $N$(1535) and
$-N$(1650) resonances.}
\label{fig:s11}
\end{figure}

\begin{figure}
\resizebox{0.9\textwidth}{!}{%
  \includegraphics{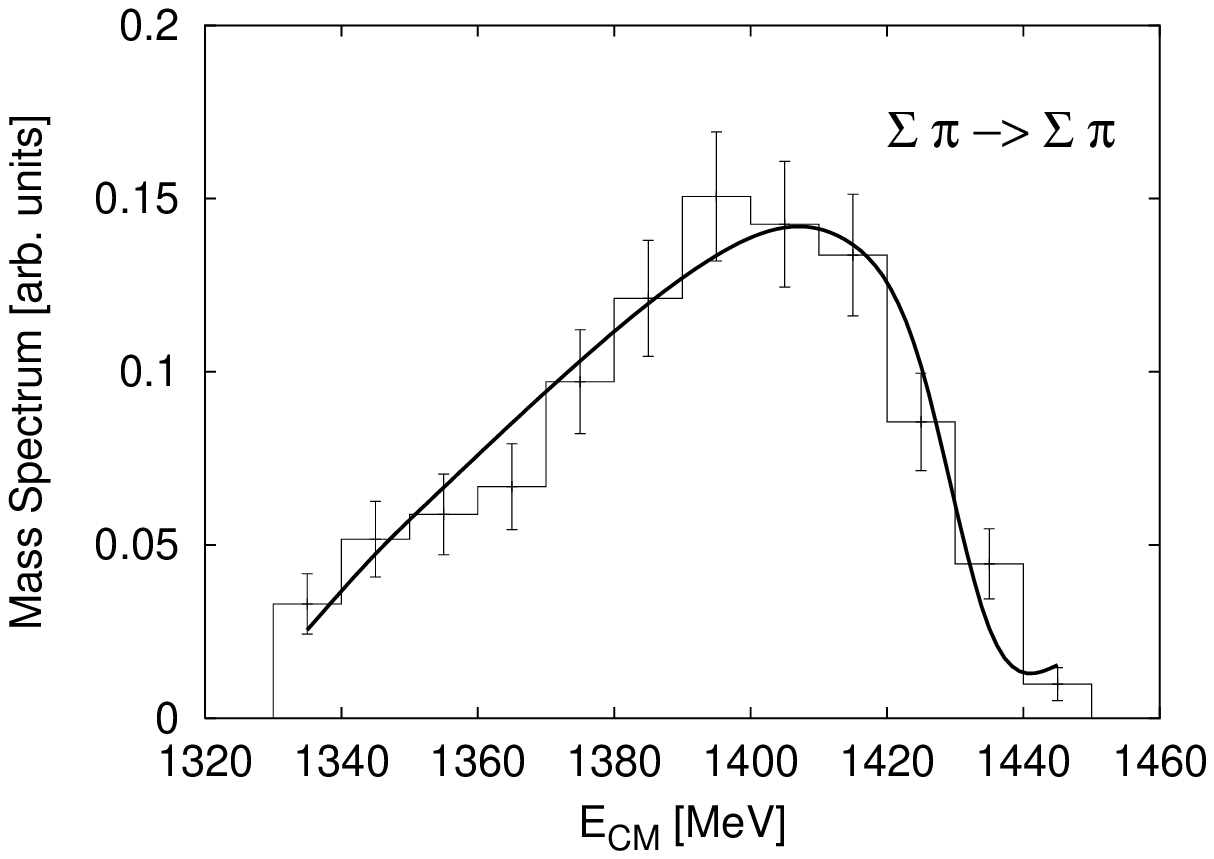}\includegraphics{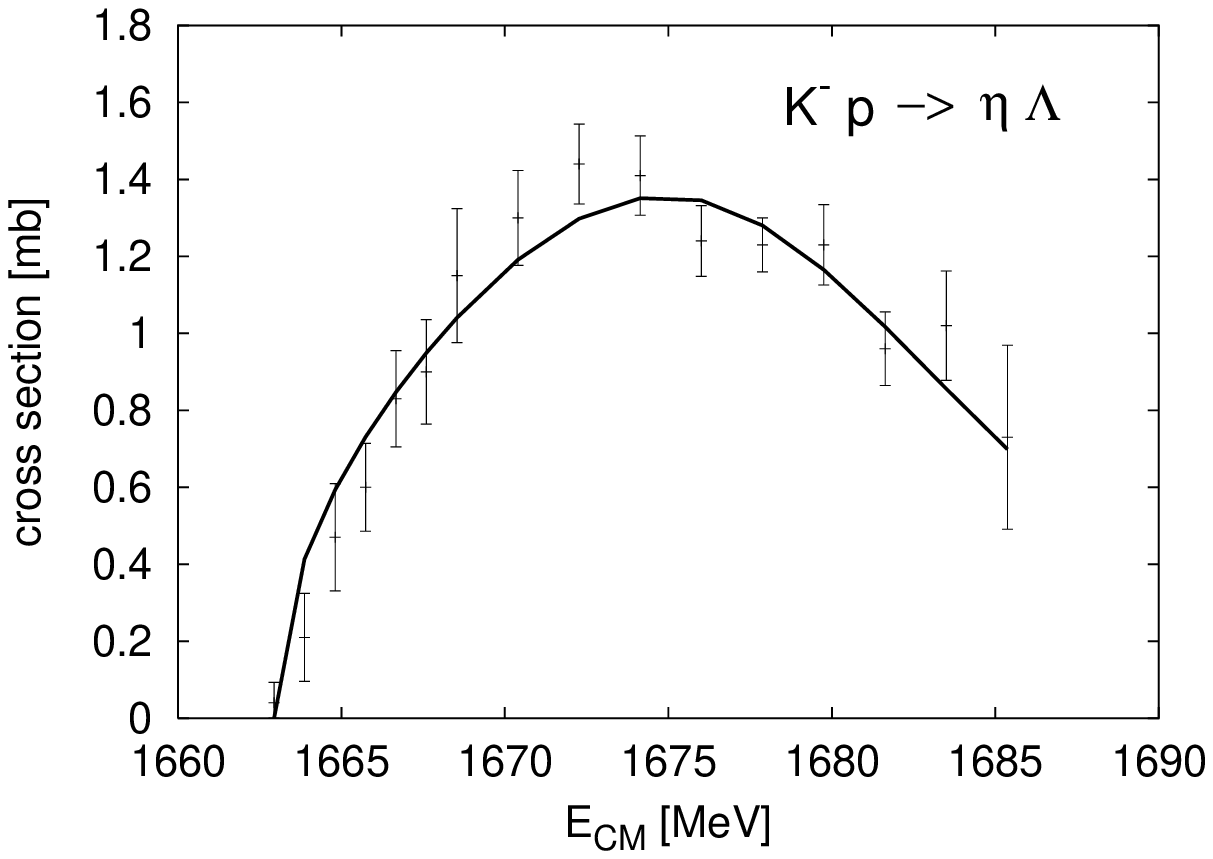}
}
\resizebox{0.9\textwidth}{!}{%
  \includegraphics{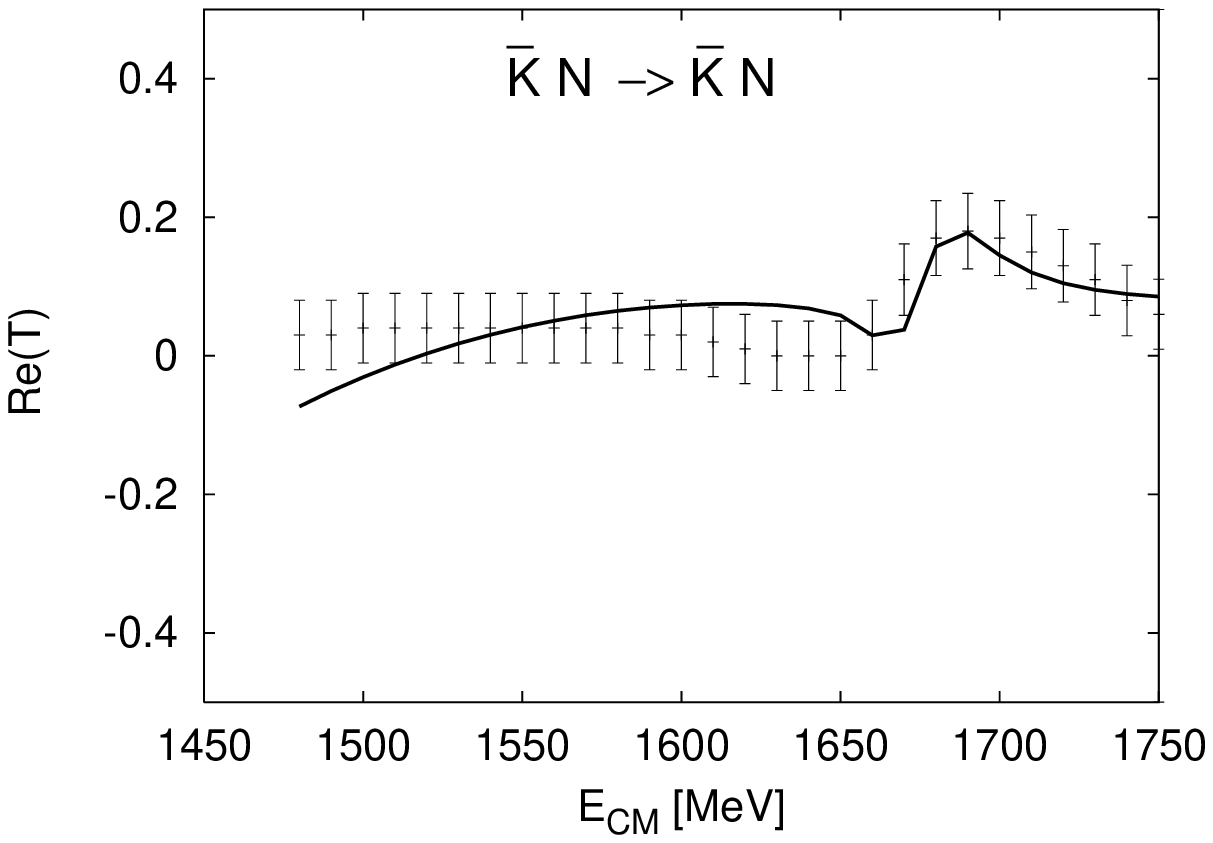}\includegraphics{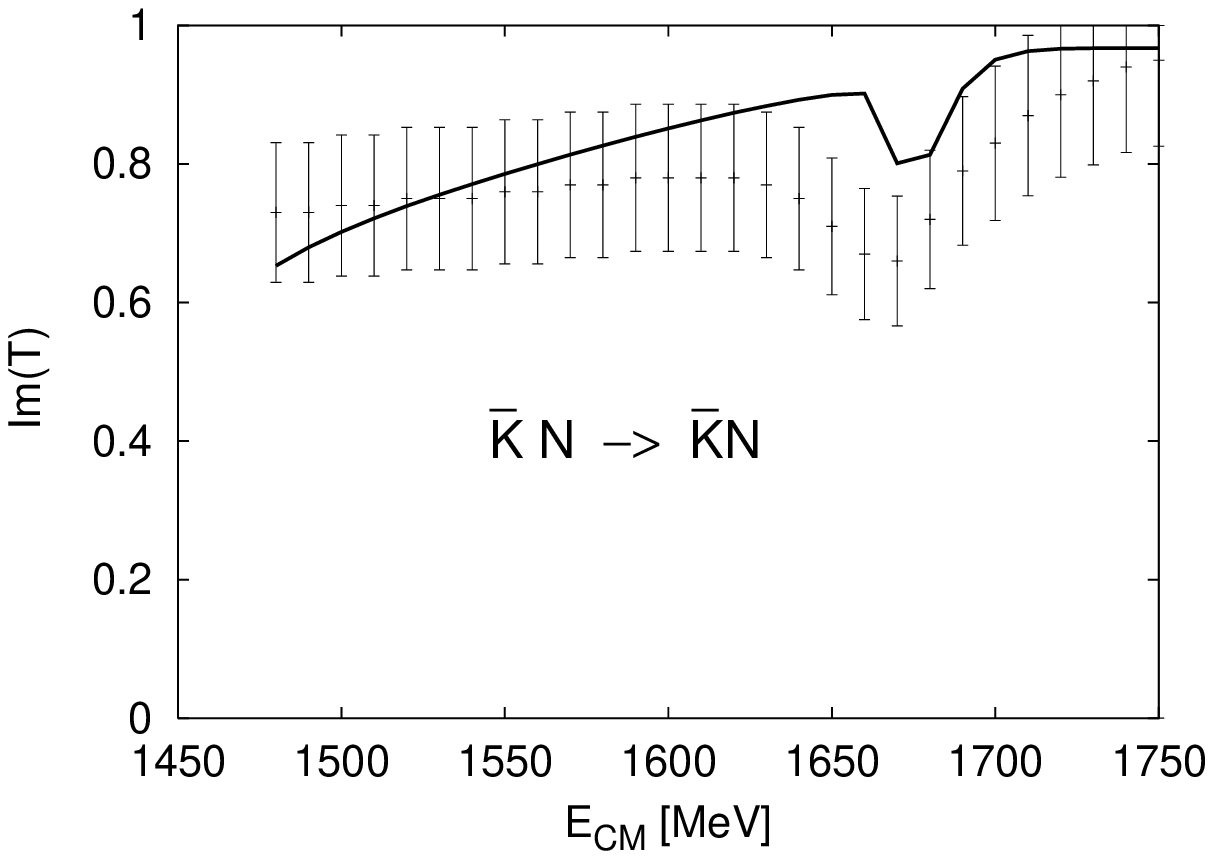}
}
\resizebox{0.9\textwidth}{!}{%
  \includegraphics{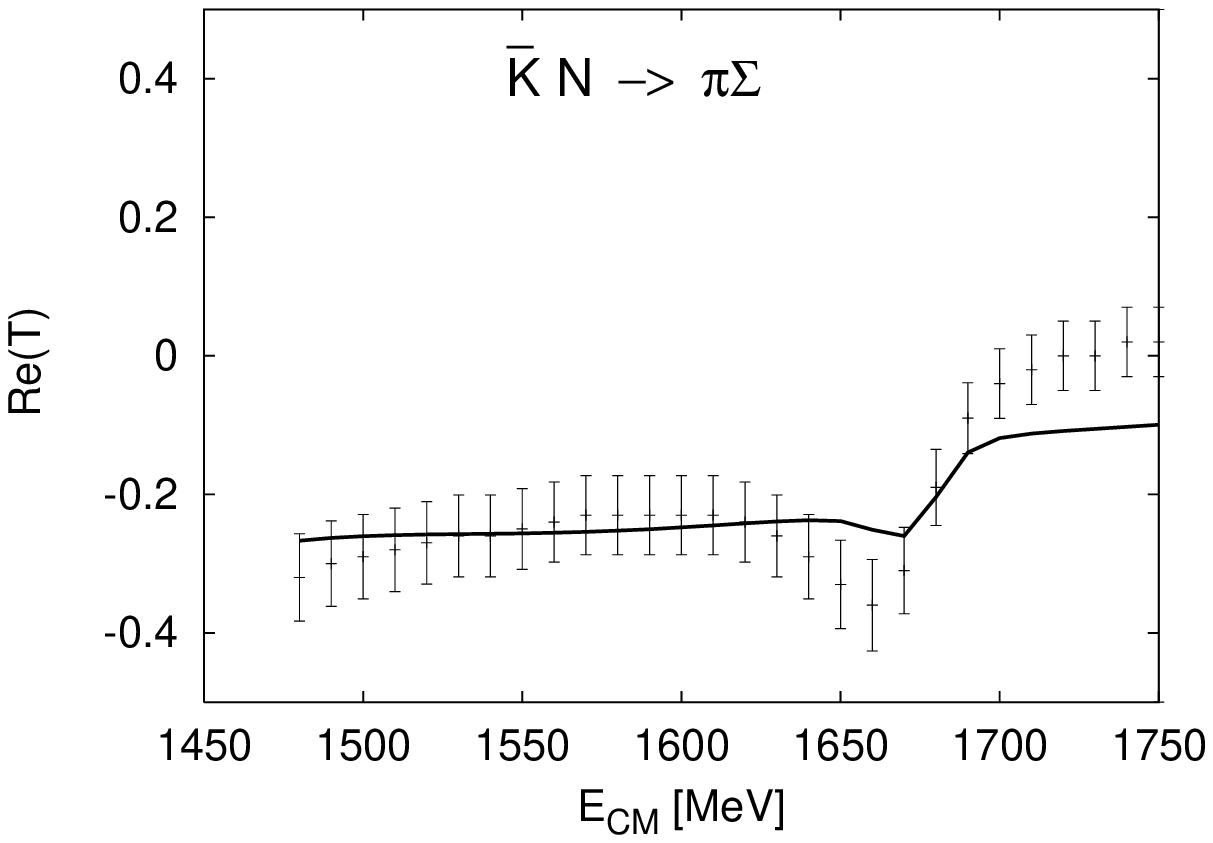}\includegraphics{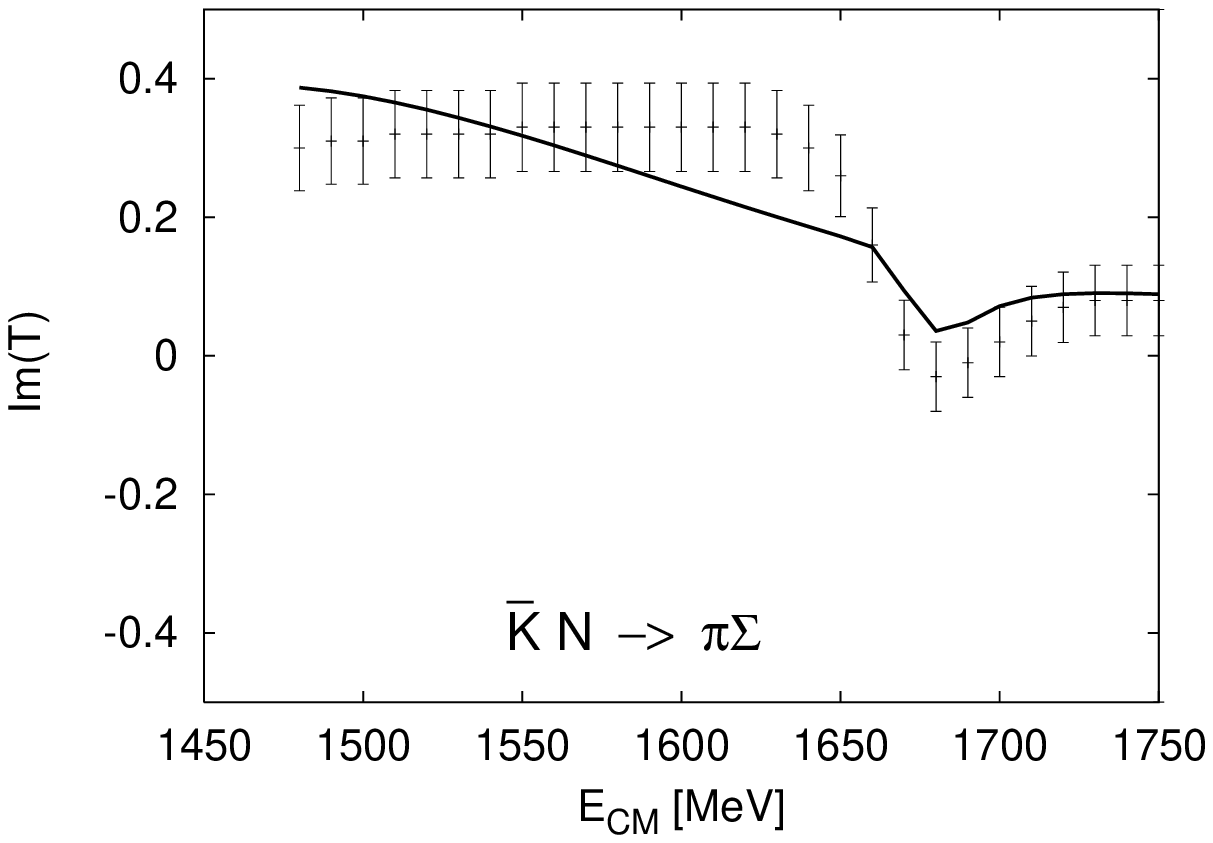}
}
\caption{Best fit results for the Bethe-Salpeter equation in the
$S=-1$, $I=0$ meson-baryon sector (solid lines). Upper panel:
$\pi\Sigma\to\pi\Sigma$ mass spectrum and $K^- p\to \eta\Lambda$ cross
section. Middle panel: The real (left panel) and imaginary (right
panel) parts of the $s-$wave $T-$matrix for elastic $\bar K N \to \bar
K N$ process in the $I=0$ isospin channel as functions of the CM
energy. Lower panel: Same as middle panel for the inelastic channel
$\bar K N \to \pi \Sigma$.}
\label{fig:s01}       
\end{figure}

For each strangeness-isospin channel, we solve the BSE 
\begin{equation} 
t_P ( k,k') = v_P ( k,k') + {\rm i} \int  \frac{d^4 q}{(2\pi)^4 }
t_P (q,k') \Delta(q) S(P-q) v_P (k,q) \label{eq:bse}
\end{equation}
where $P$ is the total four-momentum of the meson-baryon system
($s=P^2$), $k$ and $k^\prime$ are the meson incoming and outgoing
momenta, $t_P( k,k')$ is the scattering amplitude, $v_P(k,k')$ the two
particle irreducible Green's function (or {\it potential} ), and $
S(P-q)$ and $\Delta (q) $ the baryon and meson exact propagators
respectively. The above equation turns out to be a matrix one, both in
the coupled channel and Dirac spaces. For any choice of the {\it
potential} $v_P(k,k')$, the resulting scattering amplitude $t_P(
k,k')$ fulfills the coupled channel unitarity condition, discussed in
Eq.~(21) of Ref.~\refcite{JE01b}. After the needed projections, both
in Dirac space and in total angular momentum, the matrix $t_P(k,k')$
determines the $s-$wave coupled-channel matrix, $f_0^\frac12 (s)$
[$f_L^J$], with the usual normalization.  The BSE requires some input
potential and baryon and meson propagators to be solved.  From the
meson-baryon chiral Lagrangian\cite{Pich95} (see Sect. IIA of
Ref.\refcite{JE01b}), one gets at lowest order for the {\it
potential}:
\begin{equation}
v_P (k,k') = t_P^{(1)} (k,k') = \hat f^{-1} D \hat f^{-1} \left (
\slashchar{k}+\slashchar{k}' \right ) 
\end{equation}
with $\hat f$ the weak decay matrix, diagonal in the the coupled
channel space and constructed out of $f_\pi$, $f_\eta$ and $f_K$, and
$D$ the coupled-channel matrix, which is different for each
strangeness-isospin sector. The solution of the BSE with the kernel
specified above can be found in Ref.~\refcite{JE01b}.  There, the
renormalization of the obtained amplitudes is studied at length.  As a
result of the renormalization procedure, and besides the physical and
weak meson decay constants, a total amount of 12 undetermined LEC's
appear for each strangeness-isospin sector. We fit these constants to
data. In Figs.~\ref{fig:s11} and~\ref{fig:s01} we show the obtained
results. Besides, the poles in the second Riemann sheets 
provide a remarkably good description of masses, widths and branching
ratios of the $S_{11}-$ $N$(1535) and $-N$(1650) and $S_{01}-$
  $\Lambda$(1405) and $-\Lambda$(1670) resonances\cite{JE01b,prd}.

%
%


\begin{thebibliography}{99}

\bibitem{Pich95}  A. Pich, {\it Rep. Prog. Phys.} {\bf 58}  563 (1995).

\bibitem{pn00} J.R. Pel\'aez and A. G\'omez Nicola, {\it Phys. Rev.}
{\bf D62} 017502 (2000).

\bibitem{ej00c} J. Nieves and E. Ruiz Arriola, {\it hep-ph/0001013};
A. G\'omez Nicola, J. Nieves, J.R. Pel\'aez and E. Ruiz Arriola, {\it
Phys. Lett.} {\bf B486} 77 (2000).

\bibitem{JE01} 
J. Nieves and E. Ruiz Arriola, {\it Phys. Rev.} {\bf D63},  076001 (2001).

\bibitem{Meissner:1999vr}
U.~G.~Meissner and J.~A.~Oller, {\it Phys. Lett.} {\bf B500}  
263 (2001). 


\bibitem{KSW95} N. Kaiser, P.B. Siegel and W. Weise, {\it Nucl. Phys.} {\bf
A594}  325 (1995).


\bibitem{OR98}  E. Oset and A. Ramos,
{\it Nucl. Phys. }{\bf A635}  99 (1998); E. Oset, A. Ramos and C. Bennhold,
{\it Phys. Lett.} {\bf B527}  99 (2002).

\bibitem{Inou} T. Inoue, E. Oset and M.J. Vicente-Vacas,
  {\it Phys. Rev.} {\bf C65} 035204 (2002).


\bibitem{JE01b} J. Nieves and E. Ruiz Arriola, {\it Phys. Rev.} {\bf D64}
  116008 (2001).

\bibitem{LK02} M. F. M. Lutz and E. E. Kolomeitsev, {\it Nucl. Phys.} 
{\bf A700}  193 (2002). 

\bibitem{Ka01} N. Kaiser, Phys. Rev. {\bf C 64 } (2001) 045204.   

\bibitem{EJ99} 
J. Nieves and E. Ruiz Arriola, {\it Phys. Lett.} {\bf B455}  30 (1999); 
 {\it Nucl. Phys.} {\bf A679}  57 (2000).

\bibitem{prd} C. Garc\'\i a-Recio, 
J. Nieves, E. Ruiz Arriola and M. J. Vicente-Vacas, {\it hep-ph/0210311}.



\end{thebibliography}
\end{document}